\documentclass[sn-nature]{sn-jnl}% Style for submissions to Nature Portfolio journals

\usepackage{graphicx}%
\usepackage{multirow}%
\usepackage{amsmath,amssymb,amsfonts}%
\usepackage{amsthm}%
\usepackage{mathrsfs}%
\usepackage[title]{appendix}%
\usepackage{xcolor}%
\usepackage{textcomp}%
\usepackage{manyfoot}%
\usepackage{booktabs}%
\usepackage{algorithm}%
\usepackage{algorithmicx}%
\usepackage{algpseudocode}%
\usepackage{listings}%

\theoremstyle{thmstyleone}%
\theoremstyle{thmstyletwo}%

\theoremstyle{thmstylethree}%

\begin{document}

\title[Exploring near-optimal energy systems with stakeholders]{Exploring near-optimal energy systems with stakeholders: a novel approach for participatory modelling}
% Near-optimal energy transitions - Community involvement in decarbonising energy systems
% 

%TC:ignore

\author*[1]{\fnm{Oskar} \sur{Vågerö}}\email{oskar.vagero@its.uio.no}

\author[2,3]{\fnm{Koen} \spfx{van} \sur{Greevenbroek}}

\author[4,5]{\fnm{Aleksander} \sur{Grochowicz}}

\author[1]{\fnm{Maximilian} \sur{Roithner}}

\affil*[1]{\orgdiv{Department of Technology Systems}, \orgname{University of Oslo}, \orgaddress{\country{Norway}}}

\affil[2]{\orgdiv{Department of Computer Science}, \orgname{UiT The Arctic University of Norway}, \orgaddress{\country{Norway}}}

\affil[3]{\orgdiv{Earth System Science}, \orgname{Stanford University}, \orgaddress{\country{United States of America}}}

\affil[4]{\orgdiv{Department of Mathematics}, \orgname{University of Oslo}, \orgaddress{\country{Norway}}}

\affil[5]{\orgdiv{Department of Wind and Energy Systems}, \orgname{Technical University of Denmark}, \orgaddress{\country{Denmark}}}

\abstract{
Including local populations in energy transition planning can increase its legitimacy, socio-political feasibility and, in turn, climate mitigation efficacy. 
Comprehensive stakeholder involvement in participatory modelling facilitates learning; however, the choice of options and scenarios in the model can be opaque to participants.
Here we present a framework in which stakeholders engage with near-optimal modelling results holistically, and showcase our methodology in a case study of the remote Arctic settlement of Longyearbyen, Svalbard.
Participants choose across a continuum of modelling-based energy system designs in an interactive interface through which they select essentially any feasible combination of system components, and immediately see the implications of their choice.
With Longyearbyen already affected by climate change, stakeholders consistently deviate from the cost optimum and instead balance different aspects like emissions, costs, and system vulnerability.
When confronted with difficult trade-offs, participants feel better informed through our tool and successfully navigate through complex and intertwined decision-making.
}
% 150 words

\maketitle

%\noindent\textbf{Context \& scale }:

%TC:endignore

\section{Introduction}\label{sec1}
Our transitions towards more sustainable energy systems and to mitigate climate change need to accelerate, but face substantial challenges. 
Such transformations require considerable technological change, development, and innovation, but are also increasingly recognised to rely on changes in human practices \cite{perlaviciutePerspectiveHumanDimensions2021} as well as societal and citizen engagement \cite{chilversSystemicApproachMapping2021,newellNavigatingTensionsRapid2022}. 
Despite barriers of social acceptance and political feasibility \cite{wustenhagenSocialAcceptanceRenewable2007}, much research analysing future energy systems devotes surprisingly little focus to it \cite{sovacoolIntegratingSocialScience2015,trutnevyteSocietalTransformationsModels2019,susserWhyEnergyModels2022,krummModellingSocialAspects2022}. 
Energy transitions involving people in decision-making create legitimacy of the processes, enables change and can help ensure socially just outcomes \cite{michelsInnovationsDemocraticGovernance2011}. 

Participatory research in energy modelling aims to fill this gap, but currently lacks holistic and simultaneously engaging methodologies.
In this article, we introduce a multi-step framework through which policymakers, stakeholders, and modellers can learn from each other and interact to find more socially acceptable energy transitions.
We demonstrate this framework with a case study conducted in Longyearbyen, Svalbard --- an Arctic community that already faces severe ramifications of climate change \cite{sydnesLearningCrisis20152021,bradleySvalbardWinterWarming2025,lapointeClimateExtremesSvalbard2024} and is warming twice as fast as the Arctic, and seven times the global average \cite{nordliRevisitingExtendedSvalbard2020}.
Our methodology is rooted in a vast number of near-optimal realistic system designs presented in an interactive interface, allowing study participants to consider trade-offs and communicate priorities for system planning.
We introduce a modelling approach, to our knowledge the first of its kind, in which participants can interact with and decide from a vast space of feasible energy systems to design a system of their liking, while being informed of their choices' impacts.

Despite generally favourable public perceptions of renewable electricity technologies \cite{scheerPublicEvaluationElectricity2013,poortingaEuropeanAttitudesClimate2018} and transitions to low-carbon energy systems altogether in Europe, opinions vary on the favoured end-state and pathways to get there \cite{pohjolainenPublicPerceptionsClimate2018}. 
The overhaul of energy infrastructure to low-carbon systems comes with environmental \cite{giladBiodiversityLineLife2024,giladBiodiversityImpactsNorways2024} as well as social consequences \cite{goforthAirPollutionDisparities2022,sasseLowcarbonElectricitySector2023,weinandExploringTrilemmaCostefficiency2022}.
Although it is crucial to consider such (in-)direct impacts for public acceptance, e.g. querying laypeople and experts, the knowledge gaps about technologies in question can lead to misconceptions of risks and impacts \cite{volkenPerspectivesInformedCitizen2018,duboisInformedCitizenPanels2019} when surveys are used to evaluate different energy system components.
Other informative approaches such as factsheets \cite{mayerInformedPublicChoices2014}, group discussions \cite{volkenPerspectivesInformedCitizen2018}, and interactive tools \cite{mayerInformedPublicChoices2014,holzerSwissElectricitySupply2023} generate more stable preferences \cite{duboisInformedCitizenPanels2019}, create more knowledge and are additionally appreciated by participants \cite{mayerInformedPublicChoices2014,pidgeonCreatingNationalCitizen2014,volkenPerspectivesInformedCitizen2018}.
Yet, in all of the above it is challenging to represent the underlying system holistically, capturing the complex interplay of variable renewables, transmission and storage \cite{grochowiczSpatiotemporalSmoothingDynamics2024}. This motivates the usage of energy system models whose techno-economic nature gives insights about different transition pathways and which challenges exist. 
Increasingly, energy modellers engage in \emph{participatory modelling}: they involve stakeholders already at the modelling stage through scenario generation, surveys and interviews, or multi-criteria decision-making \cite{mcgookinParticipatoryMethodsEnergy2021}. 
Particular stakeholder groups may still be overlooked, as only one in six reviewed participatory modelling studies included non-academic stakeholders \cite{mcgookinParticipatoryMethodsEnergy2021}.
Nonetheless, participatory modelling can generate bottom-up knowledge of assumptions and narratives \cite{vageroCanWeOptimise2023,mcgookinParticipatoryMethodsEnergy2021}, help assess which designs would be socially accepted \cite{valenzuela-venegasRenewableSociallyAccepted2024} and facilitate mutual learning for both researchers and involved actors \cite{mcgookinParticipatoryMethodsEnergy2021}. 
With increasingly complex models, the success of such approaches hinges on transparency of assumptions and modelling tools \cite{mcgookinParticipatoryMethodsEnergy2021}, but presenting stakeholders with results based on a few pre-selected scenarios and criteria may constrain the diversity of options available, potentially based on normative preconceptions and assumptions by researchers. 
Instead, modelling frameworks in which stakeholders themselves are allowed to define metrics and explore alternative system configurations are preferred \cite{vageroCanWeOptimise2023,vageroEffectsFairAllocation2024}. 

A systematic exploration of alternative system designs is possible by describing the near-optimal feasible space of a model, also known as modelling-to-generate alternatives (MGA) \cite{decarolisUsingModelingGenerate2011}.
Looking beyond only cost-optimal modelling, MGA finds feasible solutions that are only slightly more expensive, but score better on other objectives: 
it is possible to directly search for more diverse alternatives \cite{lombardiPolicyDecisionSupport2020,pickeringDiversityOptionsEliminate2022}, classify trade-offs \cite{vangreevenbroekTradingRegionalOverall2025}, find designs that are robust to weather uncertainty \cite{grochowiczIntersectingNearoptimalSpaces2023}, or perform better on non-modelled objectives such as inequity \cite{pedersenUsingModelingAll2023}, political feasibility, energy security concerns or distributional outcomes \cite{lauMeasuringExplorationEvaluation2024,lombardiHumanintheloopMGAGenerate2025}.

MGA was originally developed to support decision-making of complex and incompletely defined problems \cite{brillModelingGenerateAlternatives1982,brillMGADecisionSupport1990} and has been used for, among other things, land-use planning \cite{brillModelingGenerateAlternatives1982}, water resource usage \cite{changUseMathematicalModels1982} and agricultural planning \cite{jeffreyNearlyOptimalLinear1992}. Already when the concept was introduced \cite{brillUseOptimizationModels1979}, MGA focused on reaching beyond the limitations of cost-efficiency and social optimality. Citizens and stakeholders may not agree on public goals and how to achieve them, which calls for new methods of using optimisation models within planning processes.
Despite the many applications of MGA, and the potential for communicating trade-offs and informing policymakers, it has largely focused on generating alternatives. Rarely have the alternatives been presented to real-world stakeholders and evaluated, even if such an application was presented already in 1990 with an interactive interface and an airline scheduling problem \cite{brillMGADecisionSupport1990} . This is particularly true for energy systems planning, where MGA has only been coupled with synthetic (i.e. assumed) stakeholder preferences based on past interviews \cite{lombardiHumanintheloopMGAGenerate2025}.

Previous participatory modelling exercises with interactive interfaces for energy systems planning have used simulation scenarios (my2050 \cite{allenCarbonReductionScenarios2013,demskiEffectsExemplarScenarios2017,pidgeonCreatingNationalCitizen2014}), technology portfolios (Riskmeter \cite{xexakisAreInteractiveWebtools2019,xexakisModelsWrongTrack2020,volkenPerspectivesInformedCitizen2018,holzerSwissElectricitySupply2023,duboisInformedCitizenPanels2019} and the Portfolio-Building Computer Decision Tool \cite{mayerInformedPublicChoices2014}) or spatial generation potential (COLLAGE \cite{flackeInteractivePlanningSupport2017}) as the underlying method for stakeholder interactions. 
All of these interactive interfaces are based on presenting simplified modelling results, which ``cannot aim to fully capture the complexity of electricity capacity planning'' (quoting the documentation of the my2050 tool \cite{departmentforenergysecurityandnetzeroCarbonCalculator2024}) such as the system interactions represented by energy system models. 
Technology portfolios may be informed by energy systems modelling, but results do not include the complementarity of technologies or detailed supply-demand balancing. 
Similarly, using spatial generation potential only considers technologies individually and annual energy generation potential. 
% The documentation of the Excel-based simulation model in the my2050 tool even state: \textit{``This is a simple model and cannot aim to fully capture the complexity of electricity capacity planning''}.
Leveraging recent methodological advancements in MGA \cite{lombardiNearoptimalEnergyPlanning2025} facilitates energy systems planning with results that are both feasible and diverse in terms of structure. 
Unlike the tools outlined above, this means that complex system interactions between variable renewable resources, storage and backup generation are taken into account. 
Moreover, this means that total system cost (both capital and operational costs) can be accounted for accurately --- a feature that is also only possible because the near-optimal solutions are computed using a model with high temporal resolution.
Here we bridge these gaps by creating a framework, based on MGA, through which near-optimal spaces describe the diversity of possible solutions which are assessed and communicated through an interactive interface. 

% What will we learn?
% How are we contributing and why does it matter? 
In this work, we investigate how participatory modelling by near-optimal solutions can inform 1) stakeholders about transition possibilities and 2) policymakers about stakeholder's priorities. 
We demonstrate the methodology with a holistic view on the move away from coal in the remote Arctic settlement of Longyearbyen, Svalbard. 
Its remote location and inability to externalise the impact of energy system infrastructure makes it overseeable and particularly suitable for stakeholder engagement, and a possible blueprint for other fossil-driven Arctic communities.
We pre-compute 56,050 feasible energy system configurations with an energy system optimisation model (PyPSA-LYB) leading to diverse outcomes for the community which are accompanied by metrics measuring different qualities of each system design.
These configurations are presented to inhabitants of Longyearbyen through a novel interactive interface which allows us to assess the technological preferences, priorities and rationale of the participants while ensuring that system requirements are met at all times.
Our implementation, which dynamically updates selectable design parameters in order to restrict the user to only feasible system designs, represents a novel and accessible way to explore and interact with the vast space of options generated by near-optimal methods. 
Despite being used in a specific geographical context, the conceptual framework and tool can be implemented in other settings. 

\section{A new modelling-based interactive planning tool}

\begin{figure}
    \centering
    \includegraphics[width=\textwidth]{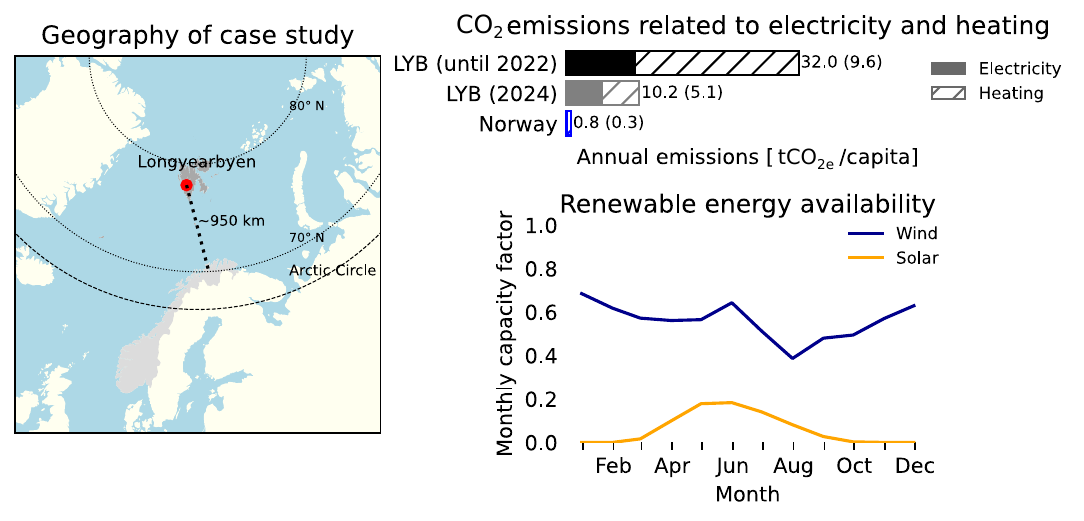}
    \caption{\textbf{Overview over Longyearbyen and its current energy system.} On the left, the map shows the geographical location of Longyearbyen on the Arctic archipelago of Svalbard as well as its distance from the north of Norway (\(\sim \)950 km) making an electricity link prohibitively expensive and unrealistic. On the top right, the per capita greenhouse gas emissions for Longyearbyen (and Norway as comparison) are plotted: for the coal-powered CHP power plant (until 2022), and the current diesel-powered CHP power plant \cite{statisticsnorwaySamfunnsforholdPaSvalbard2023,longyearbyenlokalstyreEnergiomstillingOvergangslosningBP222022}. Norwegian emissions are computed based on 2021 data on electricity consumption (\url{https://www.ssb.no/statbank/table/06913/}) and its carbon density (\url{https://www.nve.no/energi/energisystem/kraftproduksjon/hvor-kommer-stroemmen-fra/}) as well as biomass heating emissions from Arvesen et al. \cite{arvesenCoolingAerosolsChanges2018}. On the bottom right, the monthly average capacity factors for onshore wind and solar PV are plotted (for the input weather year). Note that these are subject to interannual variability and therefore just one realisation.} % Cite for Svalbard emissions SSB statistics for population and https://www.lokalstyre.no/getfile.php/5092832.2046.mlttjbuawpbatp/Orientering+overgangsløsning+BP2-2+lokalstyremøte+15112022.pdf and for Norway https://www.nve.no/nytt-fra-nve/nyheter-energi/lavt-klimagassutslipp-knyttet-til-norsk-stroemforbruk-i-2021/ and https://www.nve.no/media/13338/kvartalsrapportq4_2021.pdf and SSB for population.
    \label{fig:overview}
\end{figure}

Longyearbyen (overview in figure \ref{fig:overview}) is well-suited for energy systems research for a number of reasons. 
For one, the energy system of Longyearbyen is an isolated off-grid system, located at 78\textdegree N on Svalbard, and a mainland connection (\(\sim \)950km away) is prohibitively expensive \citep{themaAlternativerFramtidigEnergiforsyning2018}. 
The external influences on the system are therefore limited, which simplifies modelling, and at the same time makes impacts and trade-offs of different energy technologies directly apparent. 
Longyearbyen is already transitioning away from its coal-powered history, a centre of climate change research, and of extraordinary geopolitical interest \cite{osthagenMythsSvalbardGeopolitics2024} --- which keeps energy on the agenda of the local population, and is frequently discussed in local and national media. 

\begin{figure}
    \centering
    \includegraphics[width=0.7\linewidth]{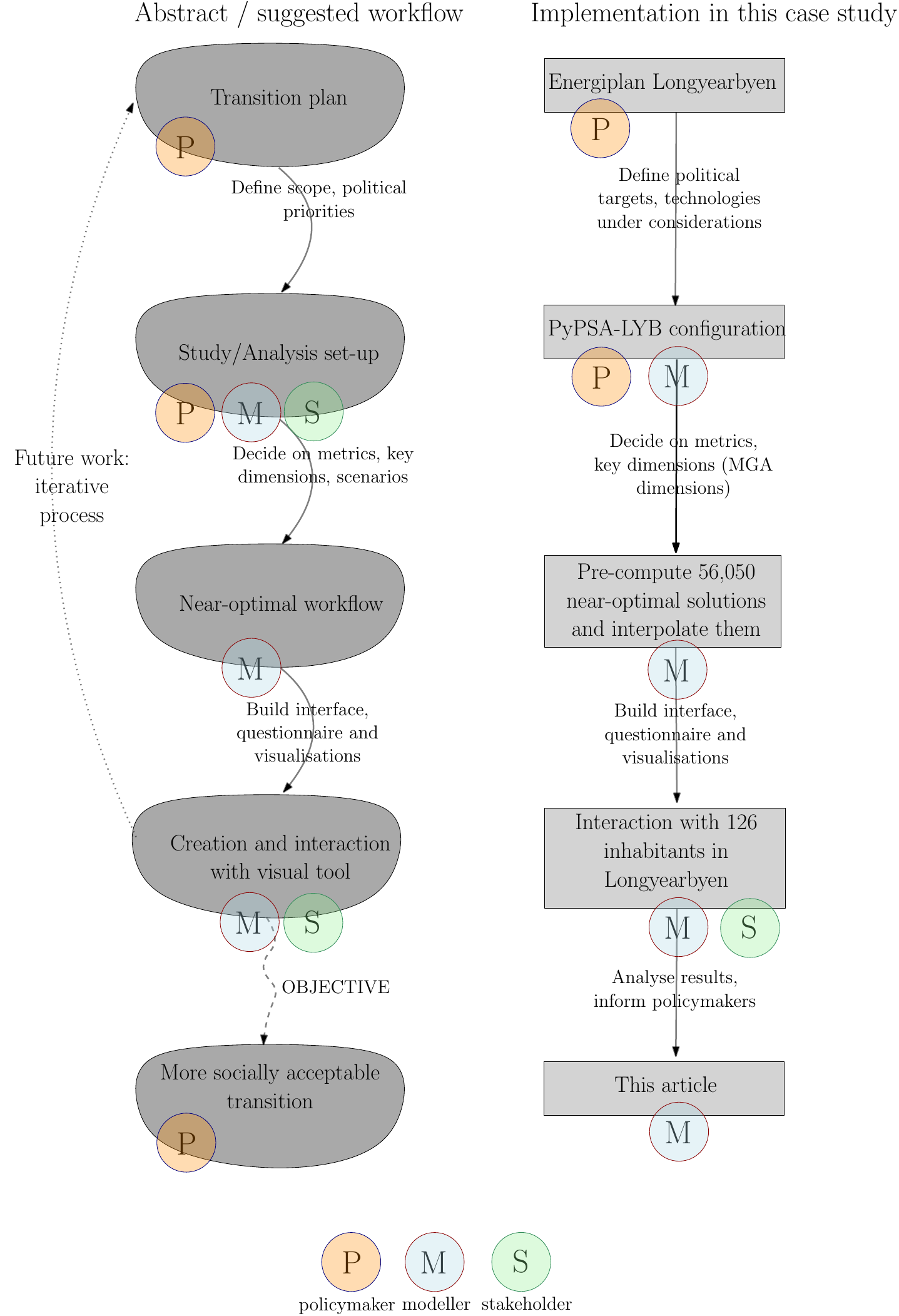}
    \caption{\textbf{Workflow for participatory modelling studies using near-optimal methods.} On the left side we suggest an abstract workflow in which policymakers, modellers, and stakeholders collaborate to develop more socially acceptable transitions. For each of these actors, the circles with ``P'', ``M'', and ``S'' indicate at which stages they are involved. On the right side, we present a concrete implementation of this workflow for our case study in Longyearbyen, starting with ``Energiplan'' as in \cite{grotteEnergiplanLongyearbyenEnergiomstilling2023}.}
    \label{fig:workflow}
\end{figure}

To engage the local community in energy planning, we developed an interactive interface, which functions as a tool for users to interact with results from an energy system optimisation model, PyPSA-LYB (see Methods).  
The scenarios underlying the model have previously been used for planning the energy transition in Longyearbyen \cite{grotteEnergiplanLongyearbyenEnergiomstilling2023} and serve as a baseline for our modelling development.
With the transition plan \cite{grotteEnergiplanLongyearbyenEnergiomstilling2023} defining the limits and broad scope of modelling assumptions, we explore near-optimal solutions to paint a realistic picture of how alternative energy system designs for Longyearbyen could look like. 
In particular, we follow Grochowicz et al. \cite{grochowiczIntersectingNearoptimalSpaces2023} to generate the near-optimal feasible space and sample 56,050 different energy system configurations (see Methods).

We engaged with the local population in public spaces between March 11 and 15, 2024 (N=126); each participant interacting with the interface individually. 
In the interface, participants explored a continuum of energy system designs, by allocating investment in different system components. 
Within the feasible bounds (pre-computed in PyPSA-LYB) and a generous cost increase (an upper limit of 125\% on top of the optimal costs), participants freely set investment levels for onshore wind power, solar power, imported green fuels, heat storage and hydrogen infrastructure (see Methods). 
Figure \ref{fig:workflow} illustrates both the generalised (left) and particular (right) workflow used.

\begin{figure}
    \centering
    \includegraphics[width=\textwidth]{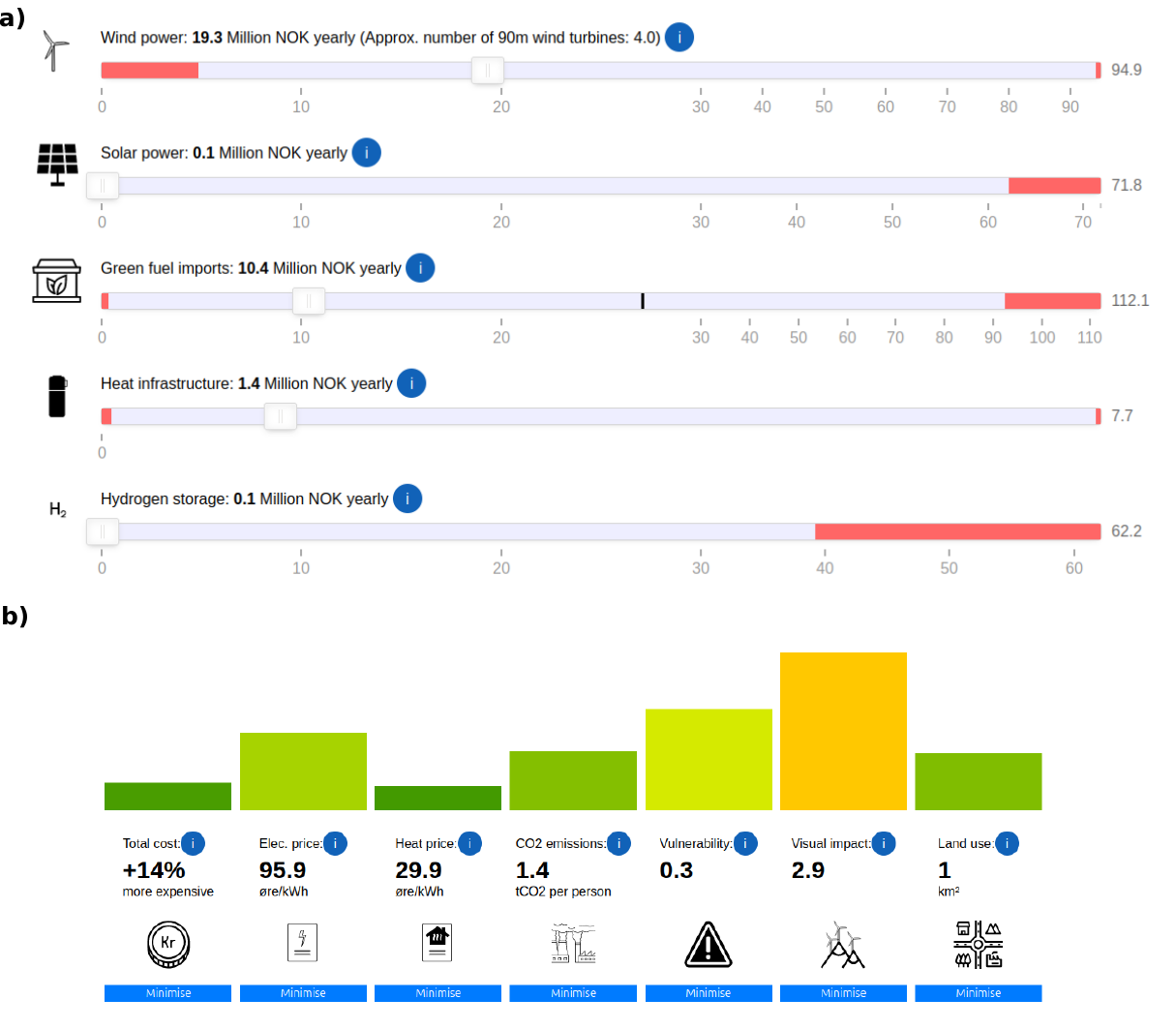}
    \caption{\textbf{Overview of the main page of the user interface.} The sliders and decision-variables which users may change are shown in \textbf{a)}, whereas \textbf{b)} shows the resulting effects of the chosen system configuration. The interface also includes a feature (buttons) which allows choosing one of the metrics to be minimised, which participants could use as a starting point. Additionally, the cost-optimal configuration is shown with a vertical line for each slider. However, it is not the starting point when opening the interface, which was chosen at random.}
    \label{fig:interface}
\end{figure}

In practice, the level of investments is controlled through sliders, as seen in figure \ref{fig:interface}a) and restricted by an updated geometry of the near-optimal space relative to the slider location. 
Apart from the investment levels, the participants were presented with key metrics of their chosen system configuration. 
These metrics are computed heuristically as needed by the interface, interpolating between pre-computed values for the 56,050 sampled configurations (see Methods). 
This feature, illustrated in figure \ref{fig:interface}b), adds relevant information and context to the decision process, and keeps track of our research question aimed at (balancing) priorities of local stakeholders. 
By complementing these quantitative values with possible qualitative preconceptions, the participants saw the impacts of their choices directly and could identify an energy system design best representing their preferences. 
After submitting their final energy system design, participants were also asked to choose up to three metrics (some chose more than three) that they prioritised for the energy system design  --- i.e. their stated priorities ---, how much additional cost they were willing to accept to improve the representation of their priorities and provide written feedback to the process. 
Complete details on the interface, as well as the rationale for choosing the particular dimensions and metrics are described in Methods. 

\section{Cost-optimality paradigms cannot meet diverse expectations and technological perspectives}

The interface outputs showcase the variety of proposed system designs and give insights about the acceptance of technologies, the importance of specific metrics as well as a partial rebuttal of cost-optimal designs. 
Through our exemplary work comprising 5\% of the population (126 participants, demographics in Methods), we can also infer specific trends and patterns for system planning in Longyearbyen.

\begin{figure}
    \centering
    \includegraphics[width=\linewidth]{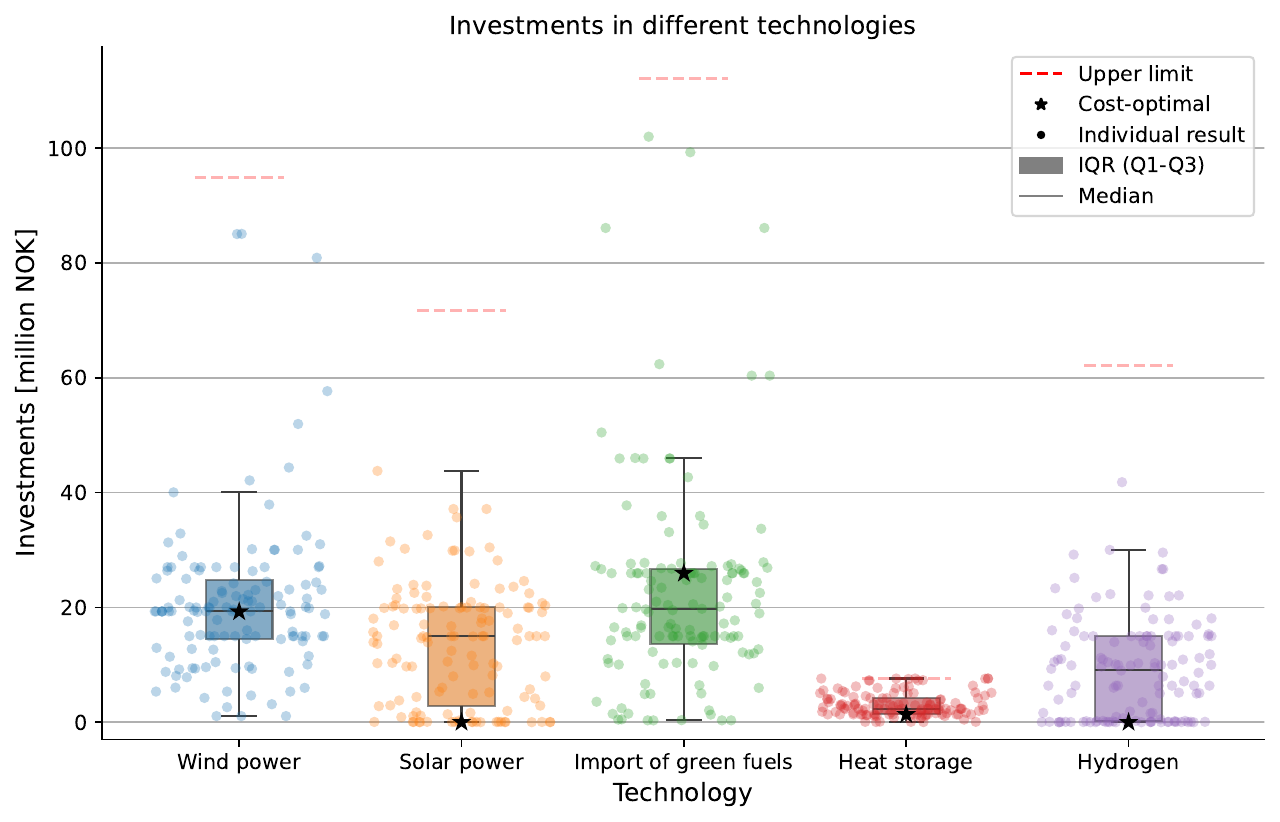}
    \caption{\textbf{Combined strip- and boxplot of submitted results across the five dimensions.} Stars represent the cost-optimal investment combination and one million NOK is approximately equivalent to 86,000 EUR or 90,500 USD in 2025. The min-max normalised median absolute deviations (MAD) quantify the distance between the median and the cost-optimum for the five dimensions, and are 0.06, 0.34, 0.11, 0.17 and 0.22, respectively. The central lines represent median values, the box limits represent the upper and lower quartiles, the whiskers extend from the box to the farthest data point lying within 1.5x the IQR. The outliers (along with all other data points) are shown in the underlying strip plot.}
    \label{fig:invest_spread}
\end{figure}

None of the technologies have consistent results with interquartile ranges (IQRs) of 10.9--36.8 percent of the decision interval. 
This is somewhat expected, as the design of the interface as well as the high slack level (125\% more expensive systems are ``near-optimal'') provide the participants with a high degree of freedom to select their preferred combination of technologies. 
In addition to the pre-computed 56,050 discrete energy system designs, the participants are also able to pick points anywhere within the feasible space, as the post-processing and interpolation within the interface (see Methods) increases the diversity of options. 

Some results such as the median for wind power investment and import of green fuels are relatively close to the cost-optimal investment. 
Solar power and hydrogen storage are outcompeted in the cost optimum, but we see that most of the participants have opted for considerable investment into these technologies (figure \ref{fig:invest_spread}). 
The participants deviate from the cost-optimal solution, and are willing to accept additional costs, if it improves other, non-monetised objectives. 

The resulting energy system designs of the participants show very high slack values (mean 81\% and median 91\%), raising questions about how well the additional cost is understood by the participants and who would bear that cost. 
This is striking given that almost half of the participants prioritised electricity prices when designing their system (more below). 
Such deviations from the cost optimum, however, are not unprecedented: indeed Fossas Tenas et al. \cite{fossastenasEmpiricalEvidenceSlack2025} found median slack levels of above 40\% in long-term transitions of European countries, with certain countries even exceeding the revealed slack levels of the participants.
It needs to be noted that when the participants were explicitly asked about acceptable additional cost the results were much lower (mean and median values of 26\% and 15\% respectively) in the post-study questionnaire.
This stark divide between revealed and stated preferences hints at the challenge to make results more tangible, but also shows that participants expect high performance on many priorities including affordability simultaneously. 

\section{Successful energy transitions rely on balancing trade-offs between local priorities}

While the aggregated results hint at the popularity of less cost-effective and less invasive technologies like solar power, heat and hydrogen storage, we can assess the participants' choices based on their stated priorities. 
From the post-study questionnaire, we can see that the most commonly prioritised metrics are emissions (69\%), followed by vulnerability (49.2\%) and electricity price (47\%). Visual impact (25.4\%), system costs (24.6\%), heat price (21.4\%) and land use (15.9\%) are not considered to be as important. 

\begin{figure}[h!]
    \centering
    \includegraphics[width=\linewidth]{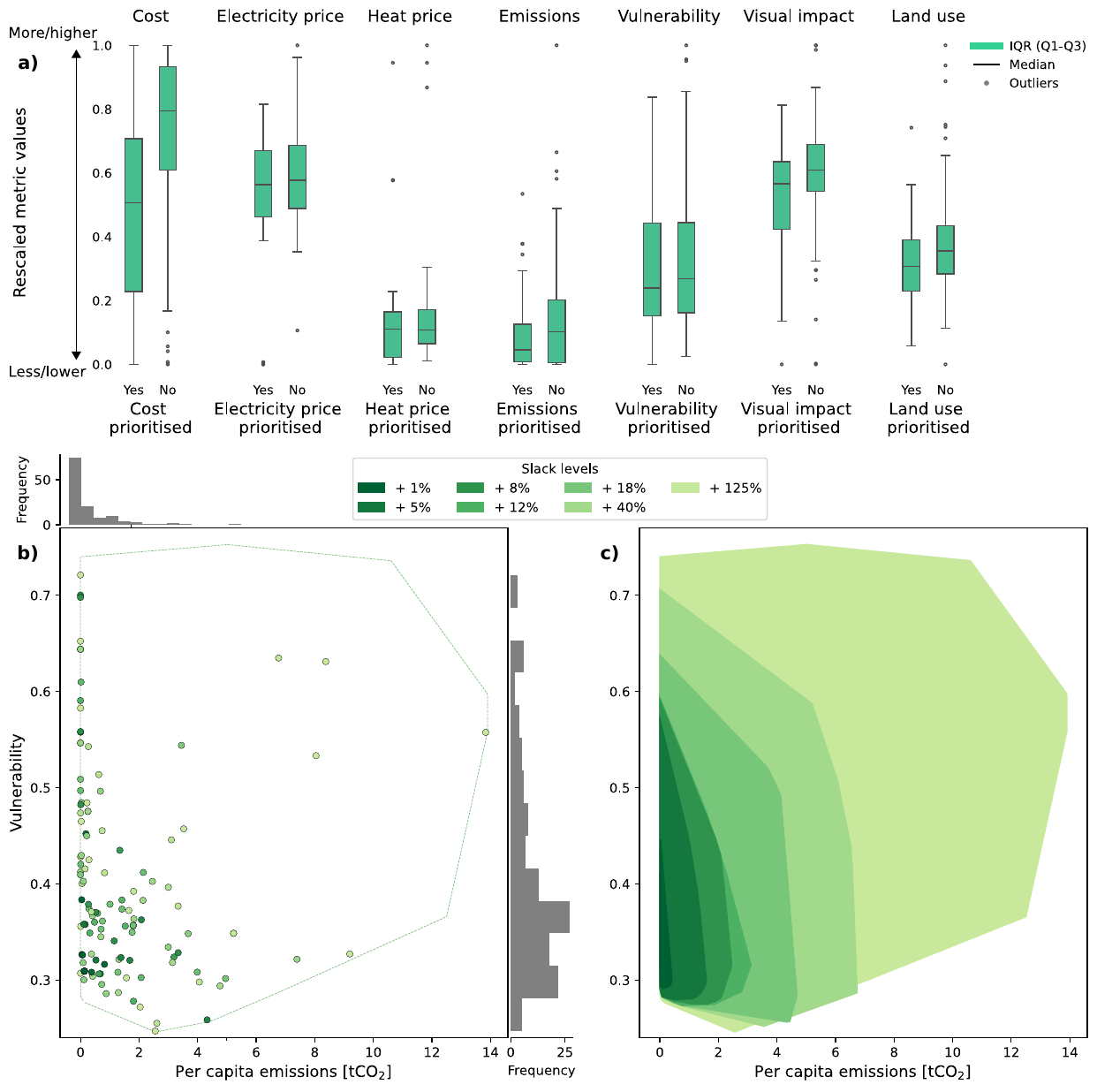}
    \caption{\textbf{Obtained values from participants depending on whether a metric was prioritised, along with associated vulnerability and emissions data}; larger difference in a) between participants who prioritised a certain metric might hint at more available options to achieve this or at higher weights of this metric. A lower value on the y-axis indicate a lower impact, which is generally desired (rescaled relative to minimum and maximum values achieved by participants). In the box plot, the central lines represent median values, the box limits represent the upper and lower quartiles, the whiskers extend from the box to the farthest data point lying within 1.5x the IQR. Outliers (fliers) are also shown. The median values as well as first and third quartiles are given in the supplementary information. Each point in b) represents a response and the according system's vulnerability and emissions. The dashed outer line represents the complete near-optimal feasible space and the colour of a point indicates the resulting slack. The histograms show the frequency of results for emissions (horizontally) and vulnerability (vertically), with 14 bins. The convex hulls for seven slack levels in c) show how the decision space changes with increasing slack.}
    \label{fig:priorities}
\end{figure}

With the flexibility and diversity of options available, participants are forced to trade off different priorities. 
Most participants (57\%) stated that they prioritised three or more metrics in their decision, introducing complexity into their decision-making. 
In fact, many participants openly acknowledged that their attempts of multi-objective planning is what they expected from policymakers too. 
Figure \ref{fig:priorities} evaluates whether participants using the tool achieved their stated priorities: we compare the obtained values of metrics and whether participants who stated to prioritise a metric when designing their system achieved a lower impact compared to those who did not prioritised it. 
Except for heat price, the group of participants who prioritised a metric, on average, identified system designs improving that metric, compared to those who did not. 
Some metrics, particularly costs, have larger differences than other.

As participants tried to balance multiple priorities, the evaluation of how well people manage to achieve their priorities is complicated (given that their individual weighing of the different metrics is unknown to us). 
Figure \ref{fig:priorities}b), illustrates how participants balanced the two most commonly prioritised metrics, vulnerability and emissions, and the resulting slack value through the shade of the point. 
Most of the participants have chosen systems with very low emissions with larger variability for vulnerability. 
Furthermore, figure \ref{fig:priorities}c) shows how the decision space of the participants grows with increasing slack values. 
While it is possible to balance these two priorities, even at low slack levels, further reductions of vulnerability come at the expense of a more expensive system. 

\begin{figure}
    \centering
    \includegraphics[width=0.7\linewidth]{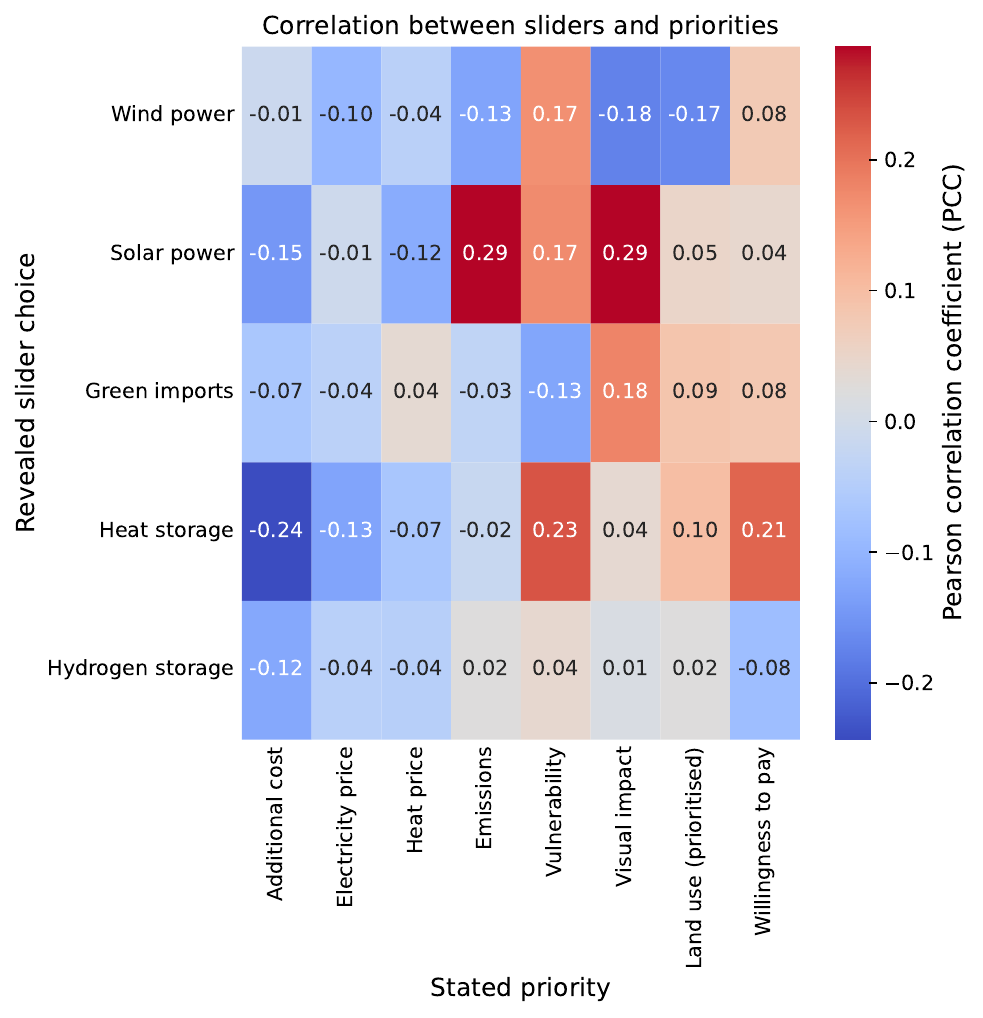}
    \caption{\textbf{Correlation matrix between revealed slider choice and stated priority of participants.} Correlations are pairwise and based on the Pearson correlation coefficient (PCC).}
    \label{fig:correlations}
\end{figure}

The correlation matrix between stated priorities and revealed slider choices in figure \ref{fig:correlations} unveils patterns between the choices of the participants and their stated priorities. 
Participants prioritising additional costs generally reduced investment (and over-sizing of the system), whereas prioritising emissions correlates with high investment in solar power, and vulnerability with additional investment into heat storage and variable renewables. 
Focusing on visual impact naturally led to reduced investment in wind power (the culprit for visual impact), and instead gravitated towards elevated investments in solar power and green imports. 
Participants who stated a high willingness-to-pay also leaned into heat storage investment. 

Since the interface was restricted to feasible system configurations, participants were unable to simply minimise all of their priorities, and instead forced to make trade-offs and balance their priorities. 
This challenged some of the participants to think about the complexity (R01, R02, R03 in table \ref{tab:feedback}), but also contributed to a feeling of being more informed. 
Furthermore, there were also participants who made the case for technologies and solutions that were not available in the interface, such as returning to a coal-powered energy supply (R04, R05), demand reducing measures (R06) and nuclear power (R07). 

\section{Discussion}\label{sec:discussion}
% 1,154 words on 22/1
We use the near-optimal feasible space of an energy system model to generate 56,050 different energy system designs for Longyearbyen, Svalbard. These system configurations were presented to 126 local citizens who explored different investments in energy system components through an interactive interface. It allowed the participants to explore trade-offs between and implications of different options. Compared to previous interactive modelling tools \cite{volkenPerspectivesInformedCitizen2018,holzerSwissElectricitySupply2023,duboisInformedCitizenPanels2019,xexakisAreInteractiveWebtools2019,xexakisModelsWrongTrack2020,flackeInteractivePlanningSupport2017,allenCarbonReductionScenarios2013,demskiEffectsExemplarScenarios2017}, applying near-optimal modelling for participatory exercises involving stakeholders allows retaining system interactions of different components as well as intermittency of renewables. 
Most critically, this approach limits the decision space to feasible system configurations. 
This prevents users from selecting systems that fail to reliably supply energy, which is of utmost importance in vulnerable communities like Longyearbyen.
Additionally, this framework opens up for discussing a diversity of options which cost-optimal scenarios overlook, reducing researcher bias, especially if stakeholders are involved at an early stage of the modelling process. 

Stated and revealed preferences in the modelling (also figure \ref{fig:invest_spread}) demonstrate that participants do not only care about the cost of their energy (system). Most participants state to balance at least three priorities, hinting at the complexity of several simultaneous trade-offs when making decisions. The common, narrow focus on cost optimality in energy modelling and policymaking is therefore potentially overlooking near-optimal solutions with higher social acceptance. The interface we present gives stakeholders the opportunity to interact with options and trade-offs, as well as to explore their priorities and how to achieve them. At the same time, it adds the system knowledge and techno-economic value of energy system modelling.

The five dimensions/sliders that the participants were adjusting were meant to represent the most relevant technologies, based on the local energy transition plan \cite{grotteEnergiplanLongyearbyenEnergiomstilling2023}:
our modelling followed the political and technological realities \cite{grotteEnergiplanLongyearbyenEnergiomstilling2023}, yet both during the process and in the post-study feedback, participants mentioned the absence of desired technologies like coal, nuclear and geothermal energy. Despite unprecedented warming and landslides in town, mentions of reviving the discontinued coal-fired thermal power plant relate to Longyearbyen's coal mining history which formed local identity and legitimised Norwegian presence on Svalbard \cite{nygaardPhaseoutsEdgeWorld2024}.
The interest in technologies that are considered infeasible (see the Methods as to why) highlights the importance of participatory studies as they reveal opinions, and help to understand why they are formed and keep society informed.
Furthermore, some participants stated a preference for reduced demand / consumption, which could indeed reduce trade-offs and repercussions of new energy infrastructure developments.
This would be an interesting dimension to include in such an interface, but how it is presented needs to be carefully considered: from a system perspective it may appear desirable, but societal and individual impacts would need to be illustrated effectively.

\subsection*{Limitations and Future Directions}

% Scope of study, involvement of stakeholders at different stages.
Participatory modelling aims to involve stakeholders, removing modeller biases in favour of exploring a breadth of socially viable alternatives.
While we demonstrate a new way for stakeholders to explore feasible system designs, large in-depth studies could involve stakeholders at additional stages in the participatory modelling workflow (figure \ref{fig:workflow}).
This could involve workshops informing participants about realistic options related to policy targets (e.g. decomissioning coal for climate mitigation), but also incorporating stakeholders in co-designing the study and interface (selecting dimensions, technologies, and metrics) directly through surveying, focus groups or other methods \cite{mcgookinAdvancingParticipatoryEnergy2024}. 
In this demonstration study, we limit our focus to the effectiveness of interface interactions.

% Effectiveness of the interface
The trade-offs around energy transitions are inherently complex, making it challenging for participants to learn from interactive interfaces in short amounts of time.
Indeed, interactive interfaces have been shown to not necessarily create more insight and understanding of the underlaying energy system dynamics than the static presentation of a few selected scenarios \cite{xexakisAreInteractiveWebtools2019} in the context of electronic surveys.
In direct feedback (table \ref{tab:feedback}), a few users were concerned about the complexity of the underlying modelling, whereas others stated that it was insightful and enjoyable and suggested making the interface available on the community council's website. 
We do not compare learning outcomes to alternatives such as a survey including fewer options, which could lead to less information overload but would be of limited utility in exploring diverse preferences.
Interactive tools may also suffer from anchoring effects \cite{demskiEffectsExemplarScenarios2017}, where the initially selected system design can influence the solution submitted by the participants.
While our study used the same (arbitrary) starting point for all participants, future work could consider using different randomly selected starting points for each participant.

% Informedness
Moreover, our study is based on relatively short interactions (typically in the range of 5--20 minutes) in public spaces, which enabled us to capture a broadly representative sample of the population (see Methods), but means that participants did not have time to learn about our modelling setup in depth.
Participants in Longyearbyen are expected to be relatively well-informed: not only is the energy system small and familiar, its transition has also been publicly discussed for many years.
The bitterly cold climate together with geopolitical tensions and impacts of climate change expose risks to various types of interruptions, and a greater degree of vulnerability than elsewhere.
It should be noted that no participant, to our knowledge, utilised or read the information sheet that we prepared and laid out. 
Instead, participants preferred and frequently made use of direct enquiries for specific explanations --- one of the supervising researchers was always at hand to answer questions.

% Specifically on the gap between stated and revealed cost slack
The discrepancy between stated willingness-to-pay (median of 15\% increase over optimality) and actual cost of submitted system designs (median of 91\% increase over optimality) highlights some of the complexities of participatory modelling.
% In contrast to common values of up to 10-15\% \cite{neumannNearoptimalFeasibleSpace2021,neumannBroadRangesInvestment2023}, we purposefully increase slack levels (125\%) to broaden the diversity of options, and instead let participants decide the slack level of their system. 
While the stated willingness-to-pay is similar to both previous studies and common MGA assumptions \cite{trutnevyteDoesCostOptimization2016, lombardiNearoptimalEnergyPlanning2025}, even the 91\% cost increase over optimality obtained from the participants lies within the range of observed deviations from the cost-optimum \cite{fossastenasEmpiricalEvidenceSlack2025}.
In effect, participants seem willing to choose expensive solutions in order to pursue non-monetary preferences, but when asked directly, are not willing to front the cost themselves.
Future studies could look into ways of making total system cost more tangible to the interface users (for instance, framing it directly as a personal cost or expected utilities bill) in order to close the gap.
In the specific case of Longyearbyen, however, some study participants may have plausibly believed that they would not personally pay for an expensive system, since the whole community is already heavily subsidised by the Norwegian government.
Thus, the large gap between stated and revealed system cost slack may be a location-specific artifact to an extent.
Further research into this topic may reveal realistic slack values for future near-optimal modelling.

In spite of the limitations outlined above, we show that participants were able to meaningfully pursue their stated objectives (figure \ref{fig:priorities}), indicating that the proposed interactive interface is effective.
Still, more in-depth engagements such as focus groups or workshops \cite{mcgookinAdvancingParticipatoryEnergy2024} could lead to even more informed responses (though at higher cost).

Growing interest \cite{vageroCanWeOptimise2023,lauMeasuringExplorationEvaluation2024} and recent advances of near-optimal methods in energy system modelling provide opportunities for further fine-tuning of the presented methodology. For example, an MGA post-processing by Lau et al. \cite{lauModellingGenerateContinuous2024} can respond ``live'' to stakeholder input when exploring the near-optimal space, potentially reducing pre-computations as well as modelling assumptions.
Lombardi and Pfenninger \cite{lombardiHumanintheloopMGAGenerate2025} propose a Human-in-the-loop (HITL) extension to conventional MGA approaches, which generates a secondary collection of near-optimal alternatives closely aligned with initially selected preferred solutions.
These, and other improvements to MGA and near-optimality in energy system models may contribute to further enhancing its usefulness for stakeholder interactions, elicitation of informed stakeholder preferences and as a tool for learning.

\subsection*{Conclusion}
For the energy transition to succeed, social and political feasibility is of utmost importance. 
Contested technologies and infrastructure risk slowing down processes and limit political will. 
Supporting legitimacy in decisions and processes, and making people's voices heard in energy transitions improves our chance of collectively limiting climate change and staying within the planetary boundaries. 
We have, with this work, demonstrated a methodology for combining near-optimality with stakeholder interactions that shows promise for illustrating and navigating difficult trade-offs in energy systems planning. 
Participants who have interacted with the methodology have clearly shown that there is more to consider than only cost-optimality, and that all preferences cannot always be met. 
Engaging with a continuum of feasible system designs via visualisation tools, such as our interactive interface, can help demonstrate the complexity in planning decisions (one of the key issues of participatory modelling) \cite{mcgookinAdvancingParticipatoryEnergy2024}, and also create meaning and a positive feeling around the challenges and dilemmas of energy transitions.

%TC:ignore
\section{Methods}\label{sec-methods}

\subsection*{Outline}

This study is based on an interface for exploring near-optimal solutions to an optimisation model for the Longyearbyen energy system.
By introducing the possibility of exploring near-optimal solutions (defined below in more detail) we can assess how different stakeholders prioritise various aspects of the energy transition and the impact on their community.
For example, one might be willing to have a slightly more expensive system to minimise a technology (e.g. solar power) or maximise other socially relevant metrics \cite{vageroCanWeOptimise2023}.
Our results are mainly based on preferences submitted by study participants through said interface, which presents the participants with sliders controlling central design parameters and a display showing the impact of design choices on a number of different metrics.

In order to power the interface, a large number of near-optimal model solutions with various total system costs are generated.
Metrics are then computed for each model solution.
In order to compute price metrics correctly, each solution was also run again in ``operational'' mode with investment variables fixed.
Study participants could then chose among these pre-computed solutions in the interactive interface.
In order to open up for additional fine-grained options and to provide a seamless user experience, the interface also allows for the selection of system designs in the convex hull of the pre-computed solutions (that is, continuous linear combinations of pre-computed solutions); metrics are interpolated for these linear combinations.

\subsection*{Modelling set-up}
The model used in this study, PyPSA-Longyearbyen or PyPSA-LYB, is an energy system optimisation model (based on the open-source modelling framework PyPSA), which optimises investment and operational decisions for a future energy system in Longyearbyen.
This model was commissioned by Longyearbyen Lokalstyret (the community council in Longyearbyen) and Svalbard Energi AS and developed by Multiconsult for the report ``Energiplan Longyearbyen'' \cite{grotteEnergiplanLongyearbyenEnergiomstilling2023} in 2023.
A number of different scenarios were explored in said report; for this study we select the model corresponding to a baseline scenario with a 2000 NOK/tCO$_2$ carbon price.
We provide a description of the model for completeness (below), but do not make any modifications to the model as obtained from Svalbard Energi.

The model covers the full energy system of Longyearbyen, including electricity, heat, as well as fully electrified transport and industrial sectors. Although the current transportation system is almost fully based on fossil fuels, many of the large organisations in Longyearbyen have implemented policies for transitioning to an electrified vehicle fleet. With additional policy measures, it is considered possible to have an electrified transportation sector by 2030 \cite{grotteEnergiplanLongyearbyenEnergiomstilling2023}. 

Longyearbyen is a small town with a mean electricity load of 3.7 MW (peak of 6.2 MW) and a mean heating load, served from a small district heating system, of 8.5 MW (peak of 18.7 MW).
Historically, both electricity and heating loads were served by a coal power plant.
The local coal mine and power plant have since been shut down and replaced with diesel generators and boilers.
Currently installed capacities (which are included in the model) consist of two diesel generators with a combined effect of 11 MW and two diesel boilers with a combined effect of 30.7 MW, as well as a 7 MWh battery.

A number of additional technologies are available in investment optimisations with zero initial capacities.
On the generation side, combined heat and power plants running on ammonia, diesel or biogas are available; see Table \ref{tab:costs} for a listing of fuel and investment costs.
The combined heat and power plants have an electricity generation efficiency of 45\%, and additionally convert 40\% of the fuel energy content to useful heat for the district heating system.
As an alternative to the conventional combined heat and power plants, the model also includes fuel cells running on the same fuels (ammonia, diesel, biogas) which have a high electricity generation efficiency (60\%) and recover less heat (20\%) --- these are also more expensive.
Boilers running on ammonia, biogas or pellets can also be invested in; these have an efficiency of 90\%.
An electric boiler with 98\% efficiency may also be invested in.
A ground heat pump with a coefficient of performance of 3 is also available.
Finally, the model includes utility solar and wind power generators at an initial capacity of 0.

A number of storage technologies are available, including battery, hydrogen and thermal storage.
Hydrogen storage tanks can be charged and discharged via electrolysis and fuel cells with efficiencies of 75\% and 50\% respectively.
Overground thermal storage (hot water tanks) is available with a maximum capacity of 100 MWh and a charge/discharge efficiency of 97.5\%.
Underground, geothermal heat storage is also available; heat can be pumped into geothermal storage at 100\% efficiency but only recovered using a heat pump with a coefficient of performance of 6 (that is, the geothermal heat pump uses 1 unit of electricity and 5 units of heat from geothermal storage to produce 6 units of heat for the district heating system).

Finally, the model allows for the recovery of low temperature waste heat from electrolysis and hydrogen fuel cells using a heat pump with a coefficient of performance of 10, with 15\% and 40\% of the input energy to electrolysis and the fuel cells recoverable as waste heat.

Coal power was excluded from the model and interface based on the political decisions that the energy supply in Longyearbyen should support and be in line with Norway's climate targets for 2030 and 2050 \cite{thenorwegianministryofjusticeandpublicsecurityMeldSt262024}. 
Nuclear is not considered techno-economically feasible, as both conventional nuclear and small modular reactors would be over-dimensioned in Longyearbyen.

\begin{table}[]
    \caption{\textbf{Operational and investment costs in the model.} Capital (investment) costs are annualised. Fuel costs are given in terms of thermal energy content using the lower heating value; the cost of diesel includes a carbon price of 2000 NOK/tCO$_2$.}
    \centering
    \begin{tabular}{lrl}
        \toprule
        Category & Cost & Unit \\ \midrule
        Diesel fuel & 1508 & NOK/MWh \\
        Ammonia fuel & 1362 & NOK/MWh \\
        Biogas fuel & 1070 & NOK/MWh \\
        Pellet fuel & 420 & NOK/MWh \\ \midrule
        Combined heat and power plant & 0.832 & MNOK/MW \\
        Fuel cell heat and power plant & 3.39 & MNOK/MW \\
        Boiler & 0.083 & MNOK/MW \\
        Electric boiler & 0.104 & MNOK/MW \\
        Ground heat pump & 3.796 & MNOK/MW \\
        Solar & 0.565 & MNOK/MW \\
        Wind & 1.151 & MNOK/MW \\ \midrule
        Battery & 0.714 & MNOK/MWh \\
        Overground heat storage & 0.027 & MNOK/MWh \\
        Geothermal storage & 100 & NOK/MWh \\
        Geothermal heat pump & 2.891 & MNOK/MW \\
        Hydrogen storage & 9110 & NOK/MWh \\
        Electrolyser & 3.79 & MNOK/MW \\
        Hydrogen fuel cell & 3.78 & MNOK/MW \\
        Hydrogen waste heat recovery & 4.819 & MNOK/MW \\
         \bottomrule
    \end{tabular}

    \label{tab:costs}
\end{table}

\subsection*{Near-optimal methodology}

A near-optimal solution (strictly speaking an $\varepsilon$-near-optimal feasible solution) describes an alternative, feasible system design whose costs are $\varepsilon$ higher than in the cost-optimal solution.
Such near-optimal solutions can be desirable because they might outperform the cost-minimal solution with regards to other qualities.
Recent research on near-optimal solutions for energy systems has shown that these system configurations can vary widely without large cost differences \cite{grochowiczIntersectingNearoptimalSpaces2023,vangreevenbroekTradingRegionalOverall2025}.
Showcasing many near-optimal solutions can help stakeholders understand the flexibility in planning of systems; if we describe near-optimal spaces geometrically, we can furthermore systematically map out trade-offs.

The space of near-optimal solutions is a convex polytope, being the feasible space of a linear program (LP), but typically consists of too many vertices in too many dimensions to be directly useful.
Thus, we follow Grochowicz et al. \cite{grochowiczIntersectingNearoptimalSpaces2023} by approximating a projection onto certain key dimensions of the near-optimal feasible space; the approximation with a limited number of vertices is obtained by repeatedly solving the original model LP with a near-optimality constraint using different, random objective functions.

In this study, the key dimensions we project down to are precisely the variables that participants are allowed to control with sliders: total investment in wind power, solar, heat storage (both overground water tank and geothermal), hydrogen storage (including storage tanks, electrolysers and fuel cells) as well as the total cost of green fuel imports (including ammonia, biogas and pellets). These five dimensions were identified as key technologies for the energy transition in Longyearbyen based on the transition plan, developed by the community council on request by the Norwegian Ministry of Petroleum and Energy (now Ministry of Energy) \cite{grotteEnergiplanLongyearbyenEnergiomstilling2023}. The transition plan evaluated different technological alternatives for electricity and heat supply in Longyearbyen, taking local needs and resource availability into account. Although the transition plan compared the alternatives with the existing diesel generators, we excluded diesel as an MGA dimension and slider (as well as reverting to the old coal power plant) based on the political decisions that Longyearbyen should support and be in line with Norway's climate targets for 2030 and 2050 \cite{thenorwegianministryofjusticeandpublicsecurityMeldSt262024,grotteEnergiplanLongyearbyenEnergiomstilling2023}. However, the existing diesel generators are still present in the model and do generate electricity. The high operational expenses means that it primarily acts as a back-up, in the absence of (cheaper) alternatives.

We chose a maximum slack level of 125\% to allow vastly different system configurations; country-level studies often use lower values instead, with one hindcasting study concluding in historic deviations from the cost-optimum of 9--24\% in the UK \cite{trutnevyteDoesCostOptimization2016}).
The method presented in Grochowicz et al. \cite{grochowiczIntersectingNearoptimalSpaces2023} leads to an approximation of the near-optimal space by a specified number of boundary vertices --- the above paper using 150 optimisations.
For the interpolation of metrics to work well, this study is based instead on a total of 56,050 optimisations at a number of different slack levels between 0 and 125\%.
This way, we obtain near-optimal solutions (for which metrics can be computed) throughout the interior and not only on the boundary of the near-optimal space.

% K: Add a brief paragraph here comparing this to other similar studies.

% TODO: It would be nice if we could find exactly which levels we used, how many optimisations per level.

\subsection*{Metrics}

For each near-optimal solution, we compute 7 different metrics which are presented in the interface:

\begin{enumerate}
    \item Total system cost slack: obtained by definition of each near-optimal solution; recall that near-optimal solutions at various cost slacks are computed in order to sample the interior of the near-optimal space.
    \item Electricity price: the shadow price of the electricity bus. Shadow prices can only be obtained through cost optimisations, so in order to compute this metric, each near-optimal solution is re-solved to cost optimality with the key design variables (wind, solar, green fuel imports, heat infrastructure, hydrogen storage) fixed.
    \item Heat price: the shadow price of the heat bus. Similar to above.
    \item CO$_2$ emissions, corresponding directly to the amount of diesel fuel used.
    \item Vulnerability ($V$): calculated as a weighted sum of dependency on weather (20\%), imports (50\%), complexity of system (10\%) and back-up heat storage (20\%):
    \begin{equation}
        V = 1 - (0.2 \cdot c + 0.5 \cdot d + 0.1 \cdot e + 0.2 \cdot f) \in [0,1],
    \end{equation} where $c$ is 1 - the share of VRE, $d$ is 1 - share of imported energy, $e$ is 1 - the share of different technologies present in the system at an installed capacity of at least 1 MW, and $f = \min\{1, heat_{pnorm}\}$ with $heat_{pnorm}$ the normalised capacity of heat storage (a value of $1$, corresponding to $4$ GWh, can cover heat demand for two weeks in the winter, and one month in the summer). In effect, a higher share of VRE, imported energy, as well as a higher complexity of the system increase vulnerability (the value of $V$), whereas higher heat storage reduces vulnerability. The components of the vulnerability metric is based on the transition plan for Longyearbyen \cite{grotteEnergiplanLongyearbyenEnergiomstilling2023}, which mention that complexity of the system, dependency on weather, reliance on import and heat storage are all important for the security of supply of the system. The weighting is based on our own conception of the hierarchy among those factors. 
    \item Visual impact: the installed wind power capacity in MW.
    \item Land use: the required use of land for energy infrastructure, see table \ref{tab:land-use}.
\end{enumerate}

\begin{table}[]
    \caption{Assumed land use requirements for technologies included in the modelling (either explicitly in the sliders or implicitly in the model). The land use requirement of different technologies is based on Chen et al. \cite{chenBalancingGHGMitigation2022} and Sasse et al. \cite{sasseRegionalImpactsElectricity2020}.}
    \centering
    \begin{tabular}{ccc}
        \toprule
        Technology & Land use (per unit) & Reference \\ 
        \midrule
        Wind power & 18000 $m^2/MW$ & Chen et al., 2022 \\
        Solar power & 50505 $m^2/MW$ & Chen et al., 2022 \\
        Bioenergy (pellets, biogas) & indirect/not local, 12.65 $m^2/MWh$ & Sasse et al., 2020 \\
        Ammonia & indirect/not local, 2.28 $m^2/MWh$ & Sasse et al., 2020 \\
        Heat storage & 1.556 $m^2/MWh$ & Chen et al., 2022 \\
        Diesel & 25 $m^2/MW$ & Chen et al., 2022 \\
        Battery storage & 6.25 $m^2/MWh$ & Chen et al., 2022 \\
        Geothermal energy & 1.556 $m^2/MWh$ & Chen et al., 2022 \\
        \bottomrule
    \end{tabular}
    
    \label{tab:land-use}
\end{table}

\subsection*{Interface and metric interpolation}

The interface presented to study participants consists of two elements (figure \ref{fig:interface}): sliders allowing the user to select and tweak an energy system design, and a display of a number of metrics associated with the selected system design (see above). 
The sliders represent the five key technologies, as explained above. Although participants were only able to change the investment into these five dimensions, the corresponding system designs include both operational decisions as well as investment decisions for other (supporting) technologies (e.g. battery storage) listed in table \ref{tab:costs}. 
These five key technologies are the primary influencing factors of the metrics in figure \ref{fig:interface}b, with the exception of CO\textsubscript{2} emissions, which is only based on the electricity generated from the diesel generators (which is not a slider).
The sliders as such allow a user to select an energy system design that matches their preferred balance of metrics, even if the selected design contains additional system components. 

Both for the sliders in the interface and in the MGA methodology described earlier, we use monetary investment (million NOK) into different technologies as the indicator, as opposed to e.g. electricity generation. 
This is partly because the five dimensions are different by nature, and do not all align well with electricity generation. 
For example, the heat infrastructure dimension is related to installing additional heat storage capacity (MWh). 
Additionally, we want to maintain a focus on the trade-off between costs and other objectives. 
Although participants clearly showed that they care for more than costs, it remains an important aspect in energy system designs.

The interactive interface is openly available in a GitHub repository (\url{https://github.com/koen-vg/near-opt-interface/}) and can be cloned and run on a personal computer. 
Alternatively, we provide an exemplary use-case for a hypothetical participant in the Supplementary Information. 

Importantly, the sliders only allow participants to select feasible system designs at most 125\% more expensive than cost-optimal.
Formally, the slider values may only encode the coordinates of points within the near-optimal space of the underlaying energy system model.
This is achieved by dynamically limiting the selectable ranges of each slider based on the currently chosen design.
Specifically, let $x_1, x_2, x_3, x_4, x_5$ represent currently selected slider values, corresponding to the key dimensions described above (wind, solar, green fuel imports, heat infrastructure and hydrogen storage), and let $\mathcal{F}_\varepsilon$ by the projection of the near-optimal space down to these five dimensions, with $\varepsilon = 125\%$.
Then $(x_1, x_2, x_3, x_4, x_5) \in \mathcal{F}_\varepsilon$.
The interface user can change a single variable at a time to select a new system design; without loss of generality, suppose this is $x_1$.
Allowable new values for $x_1$ are exactly those $x_1^*$ such that $(x_1^*, x_2, x_3, x_4, x_5) \in \mathcal{F}_\varepsilon$.
Since $\mathcal{F}_\varepsilon$ is convex, this condition is satisfied on a line segment $x_1^* \in [a, b]$, with
\begin{equation}
    a = \min x_1^* \text{ s.t. } (x_1^*, x_2, x_3, x_4, x_5) \in \mathcal{F}_\varepsilon
\end{equation}
and a corresponding condition for $b$ with a maximisation instead.
Thus, $a$ and $b$ can be computed as solutions to a simple linear program, and setting the lower and upper limits of the $x_1$ slider to $a$ and $b$ ensures that new selections stay within $\mathcal{F}_\varepsilon$.
New lower and upper limits are calculated by solving the above optimisation problems for each $x_i$ every time the user changes any of the slider values.
Regions beyond the lower and upper selectable bounds are shaded red in the interface to indicate infeasibility.

The second interface element consists of a display of a number of metrics describing characteristics of the selected energy system design.
Notably, some of the metrics (for instance, mean electricity price and CO$_2$ emissions) cannot be computed quickly from the five selectable model dimensions alone but are based on full operational model solutions.
Certainly, the metrics are not linear functions of the slider values.
This is the reason that metrics are pre-computed for a large number of model solutions spread throughout the interior and boundary of the near-optimal space, as described above.

When the user happens to exactly select a system design for which pre-computed metrics were recorded, the interface simply displays those pre-computed values.
Since slider values can be changed continuously (up to the limits of pixel and floating point precision) however, it is much more likely that the user selects a system design for which no pre-computed metrics are available.
In this case, a heuristic is used to compute approximate values for each of the metrics.
The heuristic consists of a linear interpolation among nearby pre-computed metric values.
Linear interpolation of metrics has previously been suggested by Lau et al. \cite{lauModellingGenerateContinuous2024}, who treat points in a near-optimal space as linear combinations of \emph{all} computed near-optimal model solutions.
While their method works well with a relatively small sample of near-optimal solutions, we apply a near-neighbour filter first in order to further reduce the computational complexity of the heuristic.

Specifically, let $X \subset \mathcal{F}_\varepsilon$ be the finite set of points for which we have near-optimal model solutions, and let $f \colon \mathcal{F}_\varepsilon \to \mathbb{R}$ be one of the metrics displayed in the interface.
Thus, the values of $f(x)$ for all $x \in X$ are pre-computed, whereas a fast heuristic is applied to approximate $f(x)$ for $x \in \mathcal{F}_\varepsilon \setminus X$.
For $x \in \mathcal{F}_\varepsilon \setminus X$, the heuristic works by finding a simplex with vertices $\{v_1, \dots, v_6\} \subset X$ such that $x$ is in the convex hull of $\{v_1, \dots, v_6\}$ and can be written as $x = c_1 \cdot v_1 + \dots + c_6 \cdot v_6$ with $0 \leq c_i \leq 1$.
Then $f(x)$ is approximated by $c_1 \cdot f(v_1) + \dots + c_6 \cdot f(v_6)$, essentially approximating $f$ linearly in the convex hull of $\{v_1, \dots, v_6\}$.
The simplex can be found by computing a Delaunay triangulation of $X$.
In our implementation, for computational efficiency, a Delauney triangulation is computed dynamically for each newly selected $x$ using only the 51 nearest neighbours in $X$ to $x$ in addition to all boundary vertices of the convex hull of $X$.

\subsection*{Data collection and field work}\label{subsec-datecollection}
Data was collected in Longyearbyen using the interactive interface between March 11 and 15, 2024. We approached the local population in different public spaces in Longyearbyen, at different times of the day. This include two cafés (Café Fruene and Café Huskies), the library (Longyearbyen Folkebibliotek), the University Centre in Svalbard (UNIS) and the grocery store (Svalbardbutikken). The locations were chosen based on the perceived potential for data collection and the diversity of participants that it would bring. Most participants were recruited in the grocery store, followed by the university centre. In total 126 participants went through the complete interface and submitted their results. A small number of people (\textless\ 5) did not submit their results and dropped out of the process. 

When recruiting participants, we also actively avoided temporary visitors who did not have residency in Longyearbyen (such as tourists). 

\begin{table}[h!]
\caption{Demographic characteristics of the sample, the population of Longyearbyen and Norway as a whole}\label{table:demographics}
\begin{tabular*}{0.85\linewidth}{@{\extracolsep\fill}lcccc@{\extracolsep\fill}}
\toprule
Variable name  & Full sample & Students excluded & LYB & Norway \\
\midrule
Median age                  & 32.5 & 38.0   & 35    & 40    \\
Male [\%]                   & 52.8 & 55.0   & 53.8  & 50.4  \\
Median \# of years in LYB   & 2    & 3      &       & -     \\
University degree           & 59.2 & 58.0   &       &       \\
Norwegian Citizenship [\%]  & 60   & 63     & 63    & -     \\
Median income               & 500,000 & 600,000 & 779,400 & 546,000 \\
\bottomrule
\end{tabular*}
\footnotetext{Data for Norway is from 2023 and information of their income was optional, leading to a smaller sample.}
\end{table}

The 126 participants represent Longyearbyen fairly well in several categories. There is a balance between male and female participants, different age groups and both those with and without Norwegian citizenship. Most of the participants had not stayed in Longyearbyen for a very long time, which is common in the very transient population of Longyearbyen. A few participants had been there for 10, 20 or even 30 years.

\begin{table}[h!]
\caption{Occupation of the participants, the population of Longyearbyen and Norway, classified according to Statistics Norway's Standard Industrial Classification 2007 (SIC 2007) }\label{table:occupation}
\begin{tabular*}{0.85\linewidth}{@{\extracolsep\fill}lccc@{\extracolsep\fill}}
\toprule
Occupation [\%] & Participants & LYB & Norway \textsuperscript{1} \\
\midrule
Mining & 5.0 & 4.9 & 0.2\\
Manufacturing & 2.0 & 2.0 & 9.4 \\
Construction & 4.0 & 13.5 & 9.4 \\
Wholesale and retail & 5.0 & 7.4 & 10.8 \\
Transportation and storage & 3.0 & 6.3 & 4.3 \\
Accommodation and food service & 7.0 & 16.8 & 3.4 \\
Information and communication & 6.0 & 4.5 & 4.2 \\
Real estate activities & 3.0 & 0.2 & 1.1 \\
Professional, scientific and technical activities & 16.0 & 4.8 & 5.8 \\
Administrative and support services activities & 3.0 & 9.2 & 4.4 \\
Public administration & 8.0 & 7.2 & 7.6 \\
Education & 12.0 & - & 8.2 \\
Human health and social work & 1.0 & 4.0 & 19.9\\
Arts, environment and recreation & 6.0 & 5.7 & 3.7 \\
Other service activities & 19.0 & - & \\
\bottomrule
\end{tabular*}
\raggedright 
\textsuperscript{1}Data according to SSB table 09174 \par
\end{table}

\begin{table}[h]
\caption{Exemplary qualitative feedback from the participants. They are referred to in text through their feedback ID.}\label{tab:feedback}%
\begin{tabular*}{\linewidth}{@{\extracolsep\fill}lp{11cm}@{\extracolsep\fill}}
\toprule
Feedback ID & Quote \\
\midrule
R01 & ``Good to see know how complicated it is, do want to minimise the different metrics, but it is already expensive now, and it will always be a compromise.'' \\
R02 & ``Interesting interlinked scenario. Funny to play around with different parameters and see how complex the energy mix is and consequences are.'' \\
R03 & ``Well done and gets you thinking.'' \\
R04 & ``Cannot choose desired energy solution (i.e. 100\% return to coal)'' \\
R05 & ``With respect to the current situation, a coal-fired power plant, which was already in operation, would be the most sensible economic and environmentally friendly solution. Longyearbyen is not equipped for the green transition'' \\
R06 & ``No miracle solution unfortunately, just need to reduce consumption in general and have to be ready to reduce comfort (not a lot of people are ready to do so).'' \\
R07 & ``Look at nuclear power. Safe, space-effective and durable.'' \\
\botrule
\end{tabular*}
\end{table}

\backmatter

%\bmhead{Supplementary information}

\bmhead{Acknowledgements}
% When acknowledging funding, our recommended best practice is that authors should acknowledge funders and grants on publications when the activities that contributed to that publication
O. Vågerö and A. Grochowicz acknowledge funding from Include – Research centre for socially inclusive energy transitions, funded by the Research Council of Norway, project no 295704. 

The funding institutions had no involvement in any part of the study's design, and the responsibility of the work lies with the authors.

We would like to thank Rasmus Bøckman and Torbjørn Grøtte at Svalbard Energi as well as Haakon Duus and Jonas Blomberg Ghini at Multiconsult for developing the version of PyPSA-LYB used and for providing access to the source code of the model as well as their input through discussions, before, during and after the project. 

As a final note, we would like to extend our gratitude to everyone who participated in the project as well as Svalbardbutikken, Longyearbyen Folkebibliotek, UNIS, Café Fruene and Café Huskies for facilitating the data collection. 

\bmhead{Author contributions}
All authors conceived the idea and designed the study. O.V., A.G. and K.vG. did the field work and collected the data. A.G. led the implementation of the near-optimal modelling. K.vG. led the development of the interactive interface. O.V. led the writing of the paper, with considerable input from the other authors. 

\bmhead{Data and code availability}
The source code for the interactive web interface and the code for generating the near-optimal solutions used in the interactive web interface is openly available on GitHub: interactive interface: \url{https://github.com/koen-vg/near-opt-interface/} and near-optimal workflow: \url{https://github.com/aleks-g/lyb}. 

The field data that support part of the findings of this study are not available owing to the confidentiality of participants’ identities. Longyearbyen is a small settlement, and providing individual data points may risk the identification of particular individuals, even if the data is anonymised. Aggregated data can be made available upon request. The processing of personal data has been assessed and deemed lawful and in compliance with data protection legislation by \textit{Sikt – Norwegian Agency for Shared Services in Education and Research} (reference number: 188418). The study participants were informed of the processing of personal data, the measures taken to ensure confidentiality and consented to the processing of personal data before participating.

\bmhead{Declaration of interests}
The authors declare no competing interests.

%TC:endignore

% References – as a guideline, we typically recommend up to 50.
\bibliography{pypsa-lyb.bib}% common bib file

\end{document}